\documentclass[12pt]{iopart}
\usepackage[dvips]{graphicx}

\newcommand{\mbr}{{\mathbf{r}}}

\newcommand{\bX}{{\bf X}}
\newcommand{\bx}{{\bf x}}
\newcommand{\by}{{\bf y}}
\newcommand{\bp}{{\bf p}}

\newcommand{\hp}{{\hat{\bf p}}}
\newcommand{\hx}{{\hat{\bf x}}}

\newcommand{\hy}{{\hat{\bf y}}}

\newcommand{\im}{{\mbox{i}}}

\newcommand{\be}{\begin{equation}}
\newcommand{\ee}{\end{equation}}
\newcommand{\bay}{\begin{eqnarray}}
\newcommand{\eay}{\end{eqnarray}}
\begin{document}
\title
{Multichannel Formalism for Positron-Hydrogen Scattering and Annihilation}
\author{S L Yakovlev$^{1,2}$ , C-Y Hu$^{2}$, D. Caballero$^2$
 }
\address{$^1$Department of Computational Physics, St Petersburg
University, St Petersburg, Russia 198504}
\ead{yakovlev@physics.csulb.edu}
\address{$^2$Department of Physics and Astronomy, California State University at Long Beach,
Long Beach, California 90840}
\begin{abstract}
A  problem to account for the direct electron-positron annihilation
in positron-hydrogen scattering above the positronium formation
threshold has been resolved within the time independent formalism.
The generalization of the optical theorem is derived
for the case when an absorption potential is present in the Hamiltonian.
With this theorem the annihilation cross section is fully determined by scattering
amplitudes. This allows us to separate out analytically the contribution of the positronium
formation from the overall annihilation cross section.
The rest is determined as the direct annihilation
cross section. It is done
uniformly below as well as above the positronium formation threshold.
The multichannel three-body theory for
scattering states in the presence of an imaginary absorption potential is
developed in order to compute the direct $e^+ e^-$ annihilation
amplitude. Special attention has been paid to an accurate definition of the coordinate part of the
absorption potential as the properly constructed zero-range potential, which corresponds to the
delta-function originated from the first order perturbation theory.
The calculated direct annihilation cross section below the  positronium formation
threshold is in good agreement with results of other authors.
The direct annihilation cross section computed with the formalism of the paper
shows nonsingular behavior
at the positronium formation threshold and is in  good agreement with existing data.
A number of $e^+ e^-$ direct annihilation cross
sections and positronium formation cross sections in the energy gap
between Ps$(1s)$ and H$(n=2)$ thresholds are reported.
A sharp increase in the calculated direct
annihilation cross section across the resonant energy is found for
all first S and P-wave Feshbach resonances.

\end{abstract}
\submitto{\jpb}
\maketitle
\section{Introduction}
In positron-hydrogen collision, if the energy is higher than the positronium formation threshold,
we have two genuine different asymptotic channels. The positron may go away leaving the hydrogen atom behind
in ground or in an excited state, or the positron and the electron may form a positronium leaving the
proton behind. This picture gets modified if we take into account the possibility that the positron and the
electron annihilate.
In case of rearrangement scattering, when the positron and the electron form positronium
in the Ps$(1s)$ state, the positron-electron pair annihilates mostly from this state.
The lifetime of the Ps$(1s)$ state depends on the total spin \cite{rich}.
In this process, once positronium is formed, the third particle does not affect the
annihilation.
The direct annihilation occurs without formation of the positronium. In this case the
positron interacts with the electron in the field of the proton only when all the three particles
are close to each other.
Thus, the direct annihilation is a genuine three-body process.

The standard approach treats the $e^+e^-$ annihilation on the basis
of the perturbation theory \cite{fraser, charlton}.
The annihilation cross section appears as a matrix element
of the coalescence operator between scattering states, which are the solution of the
unperturbed three-body Coulomb Hamiltonian. At higher energies, this formalism have difficulties.
At the positronium formation threshold, the calculated 
 annihilation cross section becomes infinite.
There has been a number of attempts to improve theory and to achieve
a unified  treatment of the positron annihilation
and the positronium formation \cite{mitroy, iks, gl}.
In one approach, to make the lifetime
of the $e^+e^-$ pair finite, an imaginary absorption potential is added to the positron-electron
subsystem Hamiltonian \cite{mitroy}.
"The loss of  particles", due to this potential, is then
interpreted as the $e^+e^-$ annihilation. The standard QED formula \cite{fraser, charlton}
for the positron
annihilation cross section  suggests the coordinate part of the absorption potential
to be the three-dimensional Dirac delta function depending on the relative position vector
between the positron and the electron. Subsequently, this kind of absorption potential was used
for computing the annihilation cross section above the positronium formation threshold
\cite{iks,igarashi2,yamanaka}.
The three-dimensional delta-function is too singular to be an ordinary potential
in the Schr\"odinger
equation \cite{delta-potF, delta-potB, DemkovO}. In the actual calculations in Refs.\
\cite{mitroy,iks,igarashi2,yamanaka} with the delta potential, this singularity was smoothed out.
Nevertheless, the mathematically sound formulation of using  the three-dimensional zero-range
potentials in three-body Coulombic systems
is still needed. It is done in this paper to complete the theory.

Direct annihilation cross section above the positronium formation threshold is subject of an
extensive study \cite{Laricchia}.
In Ref.\ \cite{iks,gl}
it was claimed that the direct process and the annihilation after the positronium formation are
inseparable  and therefore the notion of the direct annihilation above the threshold is meaningless.
In contrast, the direct cross section was defined and computed successfully below and above the Ps-formation
threshold in Ref.\ \cite{yamanaka}.  This calculation showed no  sign of the presence of any threshold
behavior in the energy dependence of the direct annihilation
cross section.

This paper is devoted to continue the study of the direct annihilation  within the
time independent formalism introduced in \cite{hu-yak-papp}. In  section 2 we define
the scattering solution for the $e^+-$H Schr\"odinger equation in the presence of an
absorption potential.
Subsequently,  the multichannel formalism, which is needed to determine scattering and absorption
amplitudes above the positronium formation threshold, is developed.
In section 3 we derive the generalization of the optical theorem
in the presence of an absorption potential.
This optical theorem is used to  separate out the annihilation
after the positronium formation cross section from the overall absorption
cross section. The remaining part is naturally the direct annihilation cross section.
The correct form of the zero-range absorption potential is introduced in  section 3.
Section 4 contains the results of calculations for the direct annihilation.
Section 5 concludes the paper.

Throughout the paper we use bold letters for vectors, e.g. $\bx$, and non-bold for their
magnitudes, e.g.
$x=|\bx|$. The unit vector pointing out in the direction of the vector $\by$ is denoted as
$\hy=\by/y$.
In the paper we work with wave-functions and their components of the form $\Psi(\bX,\bp)$ where
$\bX$ stands for
the configuration coordinate and $\bp$ for the momentum of the asymptotic motion. For the sake
of clarity of notations and to avoid overloading of formulae we drop very often the configuration
space coordinate from the notations of
wave-functions and their components leaving only the momentum explicitly.

\section{Three-body scattering formalism}
In this section we apply  
the three-body multichannel formalism  to $e^+-$H scattering.
\subsection{Three-body scattering in the presence of absorption potential}
We consider the three-body problem
with an additional imaginary potential which acts between the positron and the electron.
If the center of mass motion is removed, the Hamiltonian in the Jacobi coordinate system takes the form
\be
H=H^C+ \im gW_2({\bf x}_2),
\label{Hamiltonian}
\ee
\be
H^C=H_0+V^{C}_1(\bx_1)+V^{C}_2(\bx_2)+V^{C}_3(\bx_3),
\label{CoulHam}
\ee
\be
H_0=-\frac{\hbar^2}{2\mu_i}\nabla^2_{\bx_i}-
\frac{\hbar^2}{2\nu_i}\nabla^2_{\by_i}, \ \ \ V^{C}_i(\bx_i)=\frac{e_je_k}{x_i}.
\label{kineticPot}
\ee
Here we assign numbers 1,2 and 3  to the positron, the proton and the electron, respectively,
therefore the electric charges are expressed through the unit charge $e$ as $e_1=e_2=e$,
$e_3=-e$.
The Jacobi coordinates are defined in terms of particle radius-vectors $\mbr_i$ by the
standard formulae
\be
\bx_{i}=\mbr_j-\mbr_k, \ \ \ \by_{i}=\mbr_i-\frac{m_j\mbr_j+m_k\mbr_k}{m_j+m_k},
\label{Jacobiset}
\ee
and the reduced masses are given by
\be
\mu_i=\frac{m_jm_k}{m_j+m_k}, \ \ \ \nu_{i}=\frac{m_i(m_j+m_k)}{m_i+m_j+m_k}.
\label{redmass}
\ee
The potential $\im gW_{2}(\bx_2)$ acting between particles of the pair 2 
is defined such that $g$ is real and negative, and $W_2(\bx_2)$ is real and non negative.
In this case $\im gW_2$ is a complex absorbing potential.
We do not specify the coordinate dependence of $W_2$ yet, except of
requiring that $W_2$ is short-ranged.

The Schr\"odinger equation  for the positron-hydrogen scattering reads
\be
(H_0+V^C_1-E)\Psi^+=-(V^C_2+V^C_3+\im gW_{2})\Psi^{+}.
\label{Sch1}
\ee
The scattering solution is defined at real energy $E$ by the asymptotics as $y_1\to \infty$
\be
\Psi^+\propto \phi_1(\bx_1)
[e^{\im \bp_1 \cdot \by_1}+\frac{e^{\im \sqrt{E-\epsilon_1}\,y_1}}{y_1}F(p_1\hy_1,\bp_1)].
\label{asymp1}
\ee
Here $\phi_1(\bx_1)$ is the hydrogen ground state wave function with the energy $\epsilon_1$, and
the
incident momentum of the positron $\bp_1$ is related to the energy by the
condition $E=\hbar^/2\nu_1\, p^2_1+\epsilon_1$.
We note, that due to the asymptotics of the wave-function (\ref{asymp1}),
 the term $V^C_2+V^C_3$ in the right hand side of Eq.\ (\ref{Sch1}) is always confined into the region
 of the configuration space where the hydrogen wave function $\phi_1(\bx_1)$ is not negligible.
This makes the term $V^C_2+V^C_3$ short-range-type and hence the asymptotics
(\ref{asymp1}) in the 
coordinate $\by_1$ is free from the Coulomb contribution. This property holds true for all
equations we deal with in the paper.

Conventionally \cite{Messia},
the scattering amplitude $F$ can be
represented through the wave function.
It is done  by rewriting Eq.\ (\ref{Sch1}) in the integral
form and taking the
asymptotics $y_1\to \infty$.  By doing so, we get the Lippmann-Schwinger  equation (LSE)
\be
\Psi^+(\bp_1)=\Phi_1(\bp_1)+G_1(E^+)(V^C_2+V^C_3+\im g W_2)\Psi^+(\bp_1),
\label{LS}
\ee
where $E^+=E+\im 0$, $\Phi_1(\bp_1)=\phi_1(\bx_1)e^{\im \bp_1\cdot\by_1}$
is the solution to the channel Schr\"odinger equation
\be
(H_0+V^C_1-E)\Phi_1(\bp_1)=0,
\label{chanelScr}
\ee
and $G_1$ is the channel Green's function $G_1(z)=(z-H_0-V^C_1)^{-1}$.
The LSE (\ref{LS}) is the integral equation of the form
\begin{eqnarray}
\Psi^+(\bX,\bp_1)=\Phi_1(\bX,\bp_1)+
\nonumber \\
\int d\bX'\, G_1(\bX,\bX',E^+)[V^C_2(\bx'_2)+V^C_3(\bx'_3)+\im g W_2(\bx'_2)]
\Psi^+(\bX',\bp_1)
\label{LSinteg}
\end{eqnarray}
where $\bX=\{\bx_1,\by_1 \}$  and $\bx'_2$, $\bx'_3$ are supposed to be represented through
$\bx'_1,\ \by'_1$ by standard transformations of Jacobi coordinates.
The asymptotics of $\Psi^+(\bX,\bp_1)$ as $y_1\to \infty$
can easily be evaluated now from (\ref{LSinteg}) by taking the asymptotics of the Green's
function
\be
G_1(\bX,\bX',E^+)\propto  \frac{-\nu_1}{2\pi\hbar^2}\phi_1(\bx_1)
\frac{e^{\im \sqrt{E-\epsilon_1}\, y_1}}{y_1}\,
{\Phi_1}^*(\bX',\sqrt{E-\epsilon_1}\,\hy_1).
\ee
As the result, we get (\ref{asymp1}) with the following expression for the amplitude $F$
\be
F(\bp'_1,\bp_1)=\frac{-\nu_1}{2\pi\hbar^2}\langle\Phi_1(\bp'_1)|V^C_2+V^C_3+\im gW_2|
\Psi^+(\bp_1)\rangle
\label{satot}
\ee
where the matrix element stands  for the integral
\begin{eqnarray}
\langle\Phi_1(\bp'_1)|V^C_2+V^C_3+\im gW_2|\Psi^+(\bp_1)\rangle
=
\nonumber \\
\int d\bX'\,{\Phi^*_1}(\bX',\bp'_1)[V^C_2(\bx'_2)+V^C_3(\bx'_3)+\im g W_2(\bx'_2)]
\Psi^+(\bX',\bp_1).
\label{matelem}
\end{eqnarray}
The formula (\ref{satot}) suggests that the scattering amplitude can be split into the sum
of two terms
\be
F=F^0+\im gF^1,
\label{ampsplit}
\ee
where $F^0$ is exclusively due to the Coulomb interactions between the positron and the hydrogen
and $\im gF^1$ is due
to the absorption potential only. Let us note, that the immediate identification of these amplitudes
with pieces of (\ref{satot}) does not lead to the consistent form of the amplitudes $F^k$ since
the wave-function
$\Psi^+$ itself may be split into two parts similarly to (\ref{ampsplit}). Hence, the
contribution from different kinds of interactions cannot be separated on the basis of equation (\ref{satot}).
The appropriate  way is to rewrite the LSE
(\ref{LS}) in the form of distorted wave representation \cite{Messia}. To this end, let us recast
(\ref{LS}) into
\be
[I-G_1(E^+)(V^C_2+V^C_3)]\Psi^+(\bp_1)=\Phi_1+G_1(E^+)\im gW_2\Psi^{+}(\bp_1).
\label{LS1}
\ee
If the energy is below the positronium formation threshold, the inversion of the operator from
the left hand side can be performed with the help of the formulae
\be
[I-G_1(E^+)(V^C_2+V^C_3)]^{-1}\Phi_1=\Psi^{0+},
\label{Inv1}
\ee
\be
[I-G_1(z)(V^C_2+V^C_3)]^{-1}G_1(z)=G^C(z).
\label{Inv2}
\ee
As the result, the LSE (\ref{LS}) takes the  form
\be
\Psi^+(\bp_1)=\Psi^{0+}(\bp_1)+G^{C}(E^+)\im gW_2\Psi^+(\bp_1).
\label{DWLS}
\ee
The asymptotic analysis of equation (\ref{DWLS}) will give us the representations for amplitudes
$F^0$ and $F^1$.

The inhomogeneous term  $\Psi^{0+}(\bp_1)$ is the outgoing solution to the $e^+-$H scattering problem
without an absorption potential 
.
Following (\ref{Inv1}), this function is defined by the solution of the LSE
\be
\Psi^{0+}(\bp_1)=\Phi_1(\bp_1)+G_1(E^+)(V^C_2+V^C_3)\Psi^{0+}(\bp_1).
\ee
Similarly to (\ref{LS}), the asymptotics of the solution has the form
\be
\Psi^{0+}(\bp_1) \propto \phi_1(\bx_1)
[e^{\im \bp_1 \cdot \by_1}+\frac{e^{\im \sqrt{E-\epsilon_1}\,y_1}}{y_1}F^0(p_1\hy_1,\bp_1)]
\label{asymp0}
\ee
with the amplitude given by
\be
F^0(\bp'_1,\bp_1)=\frac{-\nu_1}{2\pi\hbar^2}\langle \Phi_1(\bp'_1)|V^C_2+V^C_3|
\Psi^{0+}(\bp_1)\rangle .
\label{F00}
\ee
The Green's function $G^C(z)$ in (\ref{DWLS}) is defined as
$G^C(z)=(z-H^C)^{-1}$. Its asymptotics as $y_1\to\infty$
reads
\be
G^C(\bX,\bX',E^+)\propto
\frac{-\nu_1}{2\pi\hbar^2}\phi_1(\bx_1)
\frac{e^{\im \sqrt{E-\epsilon_1}\, y_1}}{y_1}\,
{\Psi^{0-}}^*(\bX',\sqrt{E-\epsilon_1}\,\hy_1).
\label{GCasymp}
\ee
Introducing this asymptotics into (\ref{DWLS}) we get the explicit representation for
the amplitude
$F^1$  from (\ref{ampsplit})
\be
F^1(\bp'_1,\bp_1)=\frac{-\nu_1}{2\pi\hbar^2}\langle \Psi^{0-}(\bp'_1)|W_2|\Psi^{+}
(\bp_1)\rangle .
\label{F11}
\ee
In these formulae $\Psi^{0-}(\bp_1)$ is the solution of the $e^+-$H scattering problem with
incoming boundary conditions $\Psi^{0-}(\bp_1)\propto[\Psi^{0+}(-\bp_1)]^{*}$
when only  Coulomb interactions are taken into account in the Hamiltonian.
The formulae (\ref{F00}) and (\ref{F11}) provide us with the
desired representation for $F$ as the sum of two amplitudes, one of which ($F^0$)
is exclusively due to the Coulomb interactions between the positron and the hydrogen and
does not depend on the absorption potential and
the other one
($F^1$) is due to the absorption potential only.

The preceding analysis is not applicable if the energy $E$ is higher than the positronium formation threshold.
Indeed, whereas the formulae (\ref{Sch1}-\ref{ampsplit}) remain valid, the inversion  of
the operator
$I-G_1(E^+)(V^C_2+V^C_3)$ in the left hand side of (\ref{LS1}) cannot be performed
and, as the consequence, the equations (\ref{DWLS}-\ref{F11}) cannot be justified.
The formal reason is that the homogeneous equation
\be
\chi=G_{1}(E^+)(V^C_{2}+V^C_{3})\chi
\label{LShom}
\ee
now possesses  the nontrivial solution $\chi=\Psi^{02}(\bp_2)$ such that
\be
(H^C-E)\Psi^{02}(\bp_2)=0
\label{H^C-E}
\ee
with the asymptotics $\Psi^{02}(\bp_2)\propto \phi_2(\bx_2)e^{\im \bp_2\cdot\by_2}$.
The latter describes
the scattering of the proton off the  positronium ground state $\phi_2(\bx_2)$ with the
binding energy $\epsilon_2$.

This problem with LSE 
is well known
in the three-body scattering theory \cite{Tobocman, Newton, Schmid} and is the manifestation
of the general fact that
no single LSE 
specifies the three-body scattering wave function uniquely, if
the rearrangement channel is open. The resolution of the problem has been found by transforming 
LSE 
into the matrix equations for the components of the wave function. Proper
arranging of the interactions between the equations, which guarantees the uniqueness of the
solution,  leads to the equations known as the Faddeev three-body equations \cite{Newton,Schmid,Fadd}.
We adopt this formalism to our case in the
next subsection where it is proven that for the energy above the positronium formation
threshold, similar  to (\ref{ampsplit}), the amplitude
$F$ is given by the  formula $F=F^0_{11}+\im g F^1_{11}$
where the amplitudes $F^0_{11}$ and $F^{1}_{11}$
will be defined by formulae (\ref{F-0-11}) and (\ref{F-1-11}).

\subsection{Three-body scattering formalism above the positronium formation threshold}
Let us emphasize that the scattering problem with the Hamiltonian (\ref{Hamiltonian})
 always deals
with the single-arrangement channel $e^+-$H due to the presence of an absorption potential.
That means 
the solution to the Schr\"odinger equation
(\ref{Sch1}) has the single-arrangement asymptotics (\ref{asymp1})
irrespective that is the energy  below or above the Ps-formation threshold. However,
in order to specify the
amplitudes $F^0$ and $F^1$ one needs the solutions to SE 
with the Hamiltonian
$H^C$. 
For the energy above the Ps-formation threshold between
Ps($n=1$) and H($n=2$) thresholds the Scr\"odinger equation
\be
H^C\Psi^{0i}=E\Psi^{0i}
\label{Sch2}
\ee
has two kinds of solutions, which are specified  by the asymptotics
\be
\Psi^{0i}(\bp_i) \propto \phi_i(\bx_i)[e^{\im \bp_i\cdot\by_i} +
\frac{e^{\im p_iy_i}}{y_i}{f}_{ii}],
\ \ \ y_i\to \infty ,
\label{multichannelas1}
\ee
\be
\Psi^{0i}(\bp_i) \propto   \phi_k(\bx_k)\frac{e^{\im p_ky_k}}{y_k}{f}_{ki},
\ \ \ y_k\to \infty,\ \ \ k\ne i .
\label{multichannelas2}
\ee
Here the momenta $\bp_{i(k)}$ are related to the energy as
$E=\hbar^2/2\nu_{i(k)}\ p_{i(k)}^2+\epsilon_{i(k)}$
and indices $i(k)$ run over the
$\{1,2\}$ set.
The formulae (\ref{multichannelas1}, \ref{multichannelas2}) reflect the fact that
now the asymptotic
form of the wave function is different in different asymptotic arrangements.
This is exactly that property
of the three-body wave-function, which cannot be recovered by any single LSE.

In order to take into account 
the multichannel character of the scattering problem above the
rearrangement threshold we use the formalism of  Faddeev equations \cite{Fadd}.
Since the original
formalism is developed for the short range interaction, at the first stage we reformulate
the three-body Hamiltonian in such a way that Coulomb interactions are split into long-range and
short-range parts
\cite{merkuriev}
\begin{equation*}
V^C_i({\bf x}_i) = V^l_i({\bf x}_i,{\bf y}_i)+V^s_i({\bf x}_i,{\bf y}_i).
\end{equation*}
This splitting is made in the three-body configuration space by a smooth
splitting function $\zeta_i({\bf x}_i,{\bf y}_i)$
constructed such that $\zeta_i({\bf x}_i,{\bf y}_i)=1$ if
$x_i/x_0<(1+y_i/y_0)^\nu$ and
$\zeta_i({\bf x}_i,{\bf y}_i)=0$ if $x_i/x_0>(1+y_i/y_0)^\nu$ for some
$x_0>0,\, y_0>0$ and $0<\nu<1/2$.
With such a $\zeta_i$ the short- and long-range parts of the Coulomb
potentials are defined as
\begin{equation*}
V^s_i({\bf x}_i,{\bf y}_i)=\zeta_i({\bf x}_i,{\bf y}_i)V^C_i({\bf x}_i);
\ \ \
V^l_i=V^C_i-V^s_i.
\end{equation*}
The Hamiltonian (\ref{Hamiltonian}) is then transformed  into
\begin{equation*}
H=H^l+V^s_1+V^s_2+\im gW_2; \ \ \ H^l=T+V^l_1+V^l_2+V^C_3.
\end{equation*}
After this modification, the  components of the wave function $\Psi^+$ are
defined by formulae
\begin{eqnarray}
\Psi^+_1&=&(E^{+} -H^l)^{-1}V^s_1\Psi^+ ,
\label{fadd-comp1}\\
\Psi^+_2&=&(E^{+}-H^l)^{-1}(V^s_2+\im gW_2)\Psi^+ ,
\label{fadd-comp2}
\end{eqnarray}
where $E^{+}=E+\im 0$. Two components are enough  in our case. Indeed, the
potential $V^C_3$ between the positron and the proton is repulsive and does not supports bound states. Hence,
only two asymptotic arrangements are possible, which are covered by components $\Psi^+_{1,2}$.

It is straightforward to see that the sum of the
components recovers  the wave function
\begin{equation}
\Psi^+=\Psi^+_1+\Psi^+_2
\label{PsiPsi1Psi2}
\end{equation}
and the components obey the set of modified Faddeev equations (MFE)
\begin{eqnarray}
\label{mfe-1}
(E -H^l-V^s_1)\Psi^+_1&=&V^s_1\Psi^+_2 , \\
\label{mfe-2}
(E-H^l-V^s_2-\im gW_2)\Psi^+_2&=&(V^s_2+\im gW_2)\Psi^+_1.
\end{eqnarray}
The important feature of equations (\ref{mfe-1}, \ref{mfe-2}), with regard to the description
of the annihilation, is the fact that now the
two-body absorption potential $\im gW_2$ is incorporated into equations in two manners.
It appears
in the diagonal part on the left hand side
of the equation (\ref{mfe-2}) and in the coupling term on the right hand side.
The diagonal terms of the equations (\ref{mfe-1}, \ref{mfe-2}) are responsible for the asymptotic
configurations in channels 1 and  2, so the absorption potential in the left hand side of (\ref{mfe-2})
is natural
to associate with the annihilation after positronium formation.
The coupling term in (\ref{mfe-1}, \ref{mfe-2}) couples equations in the reaction volume, therefore
the absorption potential in the right hand side of (\ref{mfe-2}) is
the source for the direct annihilation. These associations will be put on the solid ground
in the next section.

As in the case of equation (\ref{Sch1}), the scattering solution to equations (\ref{mfe-1}, \ref{mfe-2})
is defined by the asymptotics as $y_1\to \infty$
\begin{eqnarray}
\label{psi-1}
\Psi^+_1(\bp_1)\propto \phi_1(\bx_1)[e^{\im \bp_1\cdot \by_1}+ \frac{e^{\im p_1 y_1}}{y_1}F_{11}],
 \\
\label{psi-2}
\Psi^+_2(\bp_1)\propto 0.
\end{eqnarray}
Certainly, for the amplitudes $F_{11}$ and $F$ from (\ref{satot}) the equality holds
true
\begin{equation*}
F_{11}=F
\end{equation*}
as it should be in view of (\ref{PsiPsi1Psi2}).
Again, as in the case of equation (\ref{Sch1}), the solution required by (\ref{psi-1}, \ref{psi-2})
is given by the integral form of MFE
(\ref{mfe-1}, \ref{mfe-2}) (IMFE)
\begin{eqnarray}
\label{IMFE-1}
\Psi^+_1(\bp_1)=\Phi^{0+}_{1}(\bp_1)+&(E^+ -H^l-V^{s}_1)^{-1}V^{s}_1\Psi^+_2(\bp_1),\\
\label{IMFE-2}
\Psi^+_2(\bp_1)=&(E^+ -H^l-V^{s}_2-\im gW_2)^{-1}(V^{s}_2+\im gW_2)\Psi^+_1(\bp_1).
\end{eqnarray}
Here $\Phi^{0+}_1$ stands for the solution to the channel Schr\"odinger
equation~$(H^l+V^{s}_1-E)\Phi^{0+}_1=0$,~which explicit form is
\be
(H_0+V_1-E)\Phi^{0+}_1=-(V^l_2+V^C_3)\Phi^{0+}_1.
\label{Chan1aseq}
\ee
Repeating  reasoning of (\ref{LS}-\ref{matelem}) we get the asymptotics
as $y_1\to \infty$
\begin{equation}
\Phi^{0+}_{1}(\bp_1)\propto \Phi_1(\bp_1) +\phi_1(\bx_1)\frac{e^{\im p_1 y_1}}{y_1}f^0_{11}
\label{f0-11}
\end{equation}
with
\be
f^0_{11}(\bp'_1,\bp_1)
=\frac{-\nu_1}{2\pi\hbar^2}\langle \Phi_1(\bp'_1)|V^{l}_2+V^C_3|\Phi^0_{1}(\bp_1)\rangle.
\label{f011}
\ee

The IMFE (\ref{IMFE-1}, \ref{IMFE-2}) are proven  to have the unique solution
below as well as above the rearrangement threshold \cite{Fadd,merkuriev}.
We use this property to remedy the shortcoming  of the LSE approach outlined above
in specifying the structure of the amplitude
$F$ above the Ps-formation threshold.

In order to proceed, it
is convenient to introduce  matrix notations
\begin{equation*}
{\bf H}(\im g)=\left[
            \begin{array}{cc}
            H^l+V^s_1   &    0\\
            0           &H^l+V^s_2+\im gW_2
            \end{array}
            \right],
\end{equation*}
\vskip 0.3cm

\begin{equation*}
{\bf V}=\left[
            \begin{array}{cc}
             0     & V^s_1\\
             V^s_2 & 0
             \end{array}
         \right],
\ \
{\bf W}=\left[
            \begin{array}{cc}
             0     & 0\\
             W_2 & 0
             \end{array}
         \right],
\ \
{\bf I}=\left[
            \begin{array}{cc}
             1     & 0\\
             0     & 1
             \end{array}
         \right].
\end{equation*}
Thus, the MFE set takes the form
\begin{equation}
[{\bf H}(\im g)-E{\bf I}]{\bf \Psi}({\bf p}_1)=-[{\bf V}+\im g{\bf W}]{\bf \Psi}({\bf p}_1),
\label{Matrix-MFE}
\end{equation}
where the vector solution is constructed from wave function components as
${\protect \bf \Psi}=(\Psi^+_1,\Psi^+_2)$. The IMFE set (\ref{IMFE-1}, \ref{IMFE-2}) in matrix
notations reads
\be
{\bf \Psi}({\bf p}_1)={\bf \Phi}^{0+}(\bp_1)+
[E^+{\bf I}-{\bf H}(\im g)]^{-1}[{\bf V}+\im g{\bf W}]{\bf \Psi}({\bf p}_1).
\label{Matrix-IMFE}
\ee
The vector of the inhomogeneous term is defined as ${\bf \Phi}^{0+}(\bp_1)=(\Phi^{0+}_1,0)$.
This equation can be reformulated in the form of distorted wave representation by similar way we
made for LSE (\ref{LS}).
The equation
(\ref{Matrix-IMFE}) can be recast into
\begin{equation*}
\{{\bf I}-[E^+{\bf I}-{\bf H}(\im g)]^{-1}{\bf V}\}{\bf \Psi}({\bf p}_1)={\bf \Phi}^{0+}(\bp_1)+
[E^+{\bf I}-{\bf H}(\im g)]^{-1}\im g{\bf W}{\bf \Psi}({\bf p}_1).
\end{equation*}
Then, by the use of the formulae
\be
\{{\bf I}-[E^+{\bf I}-{\bf H}(\im g)]^{-1}{\bf V}\}^{-1}{\bf \Phi}^{0+}(\bp_1)={\bf \Psi}^{0+}(\bp_1),
\label{Psi-0}
\ee
\be
\{{\bf I}-[z{\bf I}-{\bf H}(\im g)]^{-1}{\bf V}\}^{-1}[z{\bf I}-{\bf H}(\im g)]^{-1}=
[z{\bf I}-{\bf H}(\im g)-{\bf V}]^{-1}
\ee
we arrive at the desired distorted wave representation of the IMFE
\be
{\bf \Psi}({\bf p}_1)={\bf \Psi}^{0+}(\bp_1)+
[E^+{\bf I}-{\bf H}(\im g)-{\bf V}]^{-1}\im g{\bf W}{\bf \Psi}({\bf p}_1).
\label{DW-Matrix-IMFE}
\ee
From this equation it is seen that
\be
{\bf \Psi}({\bf p}_1)={\bf \Psi}^{0+}(\bp_1)+{\bf \Psi}^{1+}(\bp_1)
\label{Psi_101}
\ee
where ${\bf \Psi}^{1+}(\bp_1)$ stands for the second term in equation (\ref{DW-Matrix-IMFE}).
The equation (\ref{DW-Matrix-IMFE}) is the direct analog of (\ref{DWLS})
but, in contrast to that, is well defined below as well as above the Ps-formation threshold
and therefore  can be used to get the necessary
representation for the amplitude $F_{11}$.

The asymptotics of ${\bf \Psi}({\bf p}_1)$ is formed from contributions of both terms in (\ref{Psi_101}).
Let us consider the inhomogeneous term ${\bf \Psi}^{0+}(\bp_1)$. The equation for this term reads
\be
{\bf \Psi}^{0+}(\bp_1)={\bf \Phi}^{0+}(\bp_1)+
[E^+{\bf I}-{\bf H}(\im g)]^{-1}{\bf V}{\bf \Psi}^{0+}(\bp_1)
\label{IPsi-0}
\ee
or
\be
[{\bf H}(\im g)+{\bf V}]{\bf \Psi}^{0+}(\bp_1)=E{\bf I}{\bf \Psi}^{0+}(\bp_1).
\label{DPsi-0}
\ee
These equations are quite similar to (\ref{mfe-1}, \ref{mfe-2}) and (\ref{IMFE-1},
\ref{IMFE-2}) except the
coupling term ${\bf V}$, which does not contain the absorption potential. The asymptotics of the solution
to (\ref{DPsi-0}) as $y_1\to \infty$, which follows from (\ref{IPsi-0}), is similar
to (\ref{psi-1}, \ref{psi-2}) and for the components of ${\bf \Psi}^{0+}=(\Psi^{0+}_1, \Psi^{0+}_2)$
has the form
\begin{eqnarray}
\label{psi-01}
\Psi^{0+}_1(\bp_1)\propto \phi_1(\bx_1)[e^{\im \bp_1\cdot \by_1}+
\frac{e^{\im p_1 y_1}}{y_1}F^0_{11}]
\\
\label{psi-02}
\Psi^{0+}_2(\bp_1)\propto 0.
\end{eqnarray}
As it was demonstrated above, the formula for the amplitude $F^0_{11}$ should be derived by
taking asymptotics of the right hand side of (\ref{IPsi-0}) as $y_1\to \infty$. A minor
difference from what we have demonstrated above is that the
nontrivial contribution into the amplitude comes from not only  the Green's function source
term but also from  the driving term due to (\ref{f0-11}).
So that, the  amplitude $F^0_{11}$  is given by
\be
F^0_{11}(\bp'_1,\bp_1) = \frac{-\nu_1}{2\pi\hbar^2}[
\langle \Phi_1(\bp'_1)|V^{l}_2+V^C_3|\Phi^{0+}_{1}(\bp_1)\rangle +
\langle \Phi^{0+}_{1}(\bp'_1)|V^{s}_1|\Psi^{0+}_{2}(\bp_1)\rangle ] .
\label{F-0-11}
\ee
Now we
consider the ${\bf \Psi}^{1+}(\bp_1)$ term.
Its asymptotics as $y_1\to \infty$ is defined by
the Green's function ${\bf G}(z)=[z{\bf I}- {\bf H}(\im g) -{\bf V}]^{-1}$. This function is
a genuine three-body quantity and the asymptotics
of its matrix elements
as $y_i\to \infty$ may be written in the form
\be
G_{ij}(\bX,\bX',E^+)\propto
\frac{-\nu_i}{2\pi\hbar}\,\phi_i(\bx_i)\frac{e^{\im \sqrt{E-\epsilon_i}\,y_i}}{y_i}
{\Upsilon^{0-}_j}^*(\bX',\sqrt{E-\epsilon_i}\, \hy_i),
\label{Greensf}
\ee
where ${\Upsilon^{0-}_j}$ is the eigenfunction of the adjoint to the operator from the left hand side
of (\ref{DPsi-0}). This asymptotics has different character for $i=1$ and $i=2$.
In the first case $G_{1j}$ does not vanish with $y_1$ large since $E-\epsilon_1$ is real nonnegative.
If $i=2$, the Positronium binding energy $\epsilon_2$ becomes complex when the absorption potential is introduced
into the $e^+e^-$ Hamiltonian. That makes the relative momentum $p_2=\sqrt{E-\epsilon_2}$ complex, i.e.
$p_2=p^r_2+\im p^i_2$. So, the asymptotics of $G_{2j}$ vanish exponentially. Thus, only
$\Psi^{1+}_1$ component has the nontrivial asymptotics
\be
\Psi^{1+}_1(\bp_1)\propto
\phi_1(\bx_1)\frac{e^{\im p_1 y_1}}{y_1}\im g F^1_{11}
\label{Psi11}
\ee
with the amplitude $F_{11}$ given by
\be
F^1_{11}(\bp'_1,\bp_1)=\frac{-\nu_1}{2\pi\hbar^2}\langle \Upsilon^{0-}_2(\bp'_1)|W_2|
\Psi^+_1(\bp_1)\rangle.
\label{F-1-11}
\ee
Here $\Upsilon^{0-}_2$ is the second component of the solution to the adjoint equation
to (\ref{DPsi-0})
\be
[{\bf H}(-\im g)+{\bf V}^T-E{\bf I}]{\bf \Upsilon}^{0-}(\bp_1)=0
\label{DUips-0}
\ee
being defined  by the integral form 
\be
{\bf \Upsilon}^{0-}(\bp_1)={\bf \Phi}^{0-}(\bp_1)+
[E^-{\bf I}-{\bf H}(-\im g)]^{-1}{\bf V}^T{\bf \Upsilon}^{0-}(\bp_1).
\label{IUpsi-0}
\ee
In this equation $E^-=E-\im 0$, ${\bf V}^T$ means transposed matrix and
${\bf \Phi}^{0-}=(\Phi^{0-}_1,0)$ where $\Phi^{0-}_1$  is
the solution to (\ref{Chan1aseq}) with incoming boundary conditions.
The solution to (\ref{DUips-0}) in the case of $g=0$ takes the very simple form,
i.e. $\Upsilon^0_1=\Upsilon^0_2=\Psi^0$ where $\Psi^0$ is the three-body wave-function for the
pure Coulomb problem. So, $\Psi^0$ obeys the Schr\"odinger equation (\ref{Sch2})
and can be constructed from wave-function
Faddeev components (\ref{fadd-comp1}, \ref{fadd-comp2}) at $g=0$
as $\Psi^0=\Psi^{00}_1+\Psi^{00}_2$. The detailed information on this and other features of
the matrix equations as MFE and IMFE  and matrix Green's functions, which is
necessary for evaluations made above, can be found in Ref. \cite{yak-evans-hof}.

The formulae (\ref{F-0-11}, \ref{F-1-11}) determine the ingredients of the amplitude
$F_{11}$ uniformly
above as well as below the rearrangement threshold of the positronium formation. It
is possible to show by
preforming backwards transformations that below the Ps-formation threshold,  where
the representations (\ref{ampsplit}, \ref{F00}, \ref{F11}) are valid,
the equality
\be
F^0_{11}+\im g F^1_{11}= F^0 + \im gF^1=F
\label{F11-Feq}
\ee
holds true.
It is important to note, that from the analysis of the equation (\ref{IPsi-0}) it follows
\be
F^0_{11}=F^0+{\cal O}(g).
\label{F-0-11F0}
\ee
This means that the respective terms in the left and right hand sides of the first equation
in the chain (\ref{F11-Feq})
are not identical. In fact the amplitude $F^0_{11}$
takes into account not only the Coulomb interaction between the positron and the hydrogen, as $F^0$
does, but also the
possibility of $e^+e^-$ annihilation after the positronium formation. That is due to the presence
of the absorption potential in the diagonal part of the equations (\ref{DPsi-0}).
This means that
below the
positronium formation
threshold  the annihilation after virtual formation of the positronium is
incorporated into the
$F^0_{11}$ amplitude and consequently $F^1_{11}$ is the pure direct annihilation amplitude.
It is worth mentioning again, that the formulae
(\ref{F-0-11}, \ref{F-1-11}) determine the amplitudes uniformly  below as well as above
the positronium formation threshold, whereas it is not true for representations (\ref{F00}, \ref{F11}).
They are valid
only below the positronium formation threshold.

\section{Optical theorem in the presence of absorption and annihilation cross section
\label{OT}}
\subsection{Optical theorem\label{OT1}}
The standard optical theorem for Hermitian Hamiltonians is nothing but the
manifestation of
the flux conservation, what is equivalent to the unitarity of the S-matrix.
The absorption potential
breaks the Hermiticity   and the scattering is not unitary. The lack of unitarity is the
 measure of how much of the
flux is absorbed and in the case of the annihilation is the way to determine the
annihilation cross section.
There is extensive literature on the optical theorem but 
\cite{Messia}
is  the most suitable
for our purpose. Following this approach by multiplying the SE (\ref{Sch1}) by complex
conjugate
wave function ${\Psi^+}^*$ and subtracting the complex conjugated SE multiplied by $\Psi^+$
we arrive at
the equality
\be
{\Psi^+}^*H_0\Psi^+ -\Psi^+H_0{\Psi^+}^*=-2\im gW_2|\Psi^+|^2.
\label{nonHerm}
\ee
Integrating over the domain $\Omega_{R}=\{y_1\le R\}$, using the Green's formula and taking the
limit as $R\to \infty$, we get the following result  
\be
-2\im g \int d\bx_1 d\by_1\, W_2|\Psi^+|^2=\frac{\hbar^2}{2\nu_1}
\lim_{R\to \infty}\int R^2 d\hy_1\int d\bx_1\, \{\Psi^+,{\Psi^+}^*\},
\label{WronskForm}
\ee
which represents the balance of the flux.
Here the Wronskian $\{\Psi^+,{\Psi^+}^*\}= \Psi^+\partial_{y_1}{\Psi^+}^*-
{\Psi^+}^*\partial_{y_1}{\Psi^+}$
has to be taken at the condition $y_1=R$. Then, using the asymptotic form of $\Psi^+$
given in (\ref{asymp1}),
the normalization of the ground-state wave functions $\phi_1(\bx_1)$  and the weak asymptotics
of the plain wave (see for example \cite{Messia-was})
\begin{equation*}
e^{\im \bp\cdot \by}\propto \frac{2\pi}{\im py}\left[
-\delta(\hp+\hy)e^{-\im py}+\delta(\hp-\hy)e^{\im py}\right]
\end{equation*}
we finally arrive at the optical theorem in the presence of absorption
\be
\frac{2\nu_1(-g)}{\hbar^2p_1}\int d\bx_1 d\by_1\, W_2|\Psi^+|^2=
\frac{4\pi}{p_1}\Im\mbox{m} F(\bp_1,\bp_1) -\int d\hy\, |F(p_1\hy,\bp_1)|^2.
\label{OptTh}
\ee
The positive quantity in the left hand side of (\ref{OptTh}) is the absorption
cross section which determines in our case the overall $e^+e^-$ annihilation cross section due to the direct
process as well as due to the annihilation after (virtual if $E<\epsilon_2$ or actual if $E>\epsilon_2$)
positronium formation
\be
\sigma^a= \frac{2\nu_1(-g)}{\hbar^2p_1}\int d\bx_1 d\by_1\, W_2|\Psi^+|^2.
\label{sigma_a}
\ee
So that, the overall annihilation cross section can be computed either by the integral
(\ref{sigma_a})
or by the expression in the right hand side of (\ref{OptTh}), if the total amplitude $F$
is in possession.

In order to go beyond the standard formulation (\ref{OptTh}, \ref{sigma_a}) one needs to
use the detailed
structure of the amplitude $F$. In our case it is the representation (\ref{F11-Feq})
\begin{equation*}
F=F^0_{11}+\im gF^1_{11}
\end{equation*}
which leads to the following form of (\ref{OptTh})
\begin{eqnarray}
\sigma^a= \sigma^a_2 +\sigma^a_1
\label{sigma21}\\
\sigma^a_2= \frac{4\pi}{p_1}\Im\mbox{m}F^0_{11}(\bp_1,\bp_1)-\int d\hy\, |F^0_{11}(p_1\hy,\bp_1)|^2
\label{sigma2} \\
\sigma^a_1=
\frac{4\pi}{p_1}(-g)\Re\mbox{e} F^1_{11}(\bp_1,\bp_1)-
\nonumber\\
2(-g)\int d\hy\, \Im\mbox{m}F^0_{11}(p_1\hy,\bp_1){F^1_{11}}^*(p_1\hy,\bp_1)
-g^2 \int d\hy\, |F^1_{11}(p_1\hy,\bp_1)|^2.
\label{sigma1}
\end{eqnarray}
The quantities $\sigma^a_1$  and $\sigma^a_2$ have  meaning of annihilation cross sections for
the direct process and the process of the annihilation after the positronium formation.
To make this statement sound,
let us show that $\sigma^a_2$ is the cross section of the after the
positronium formation annihilation.
Thus, the remaining part of $\sigma^a$, what is $\sigma^a_1$, should be interpreted as
the direct annihilation cross section.

It is apparent, that the equation (\ref{sigma2})
is the optical theorem formulated for the equation
(\ref{DPsi-0}). 
Indeed, multiplying
the equation (\ref{DPsi-0})  by ${\bf \Upsilon}^{+0}$ from the left and subtracting
the equation for
${\bf \Upsilon}^{+0}$
\be
[{\bf H}(\im g)+{\bf V}^T-E{\bf I}]{\bf \Upsilon}^{0+}(\bp_1)=0
\label{DUips+0}
\ee
multiplied by ${\bf \Psi}^{0+}$ from the right and making obvious cancelations we get
\begin{equation*}
\langle {\bf \Upsilon}^{0+},H_0{\bf I}{\bf \Psi}^{0+}\rangle-
\langle  H_0{\bf I}{\bf \Upsilon}^{0+},{\bf \Psi}^{0+}\rangle=
2\im g \langle {\bf \Upsilon}^{0+}, {\bf D}{\bf \Psi}^{0+}\rangle.
\end{equation*}
Here ${\bf D}$ is a diagonal matrix $\mbox{diag}\{0,W_2\}$ and $\langle .,.\rangle$
is the scalar product in
the two dimensional complex space ${\bf C}^2$ of wave-function components.
Repeating argumentations which led us to the formula (\ref{OptTh}), we arrive at the equality
\begin{eqnarray}
\frac{2\nu_1(-g)}{\hbar^2 p_1}\int d\bx_1d\by_1\, {\Upsilon^{0+}_2}^*(\bp_1)W_2\Psi^{0+}_2(\bp_1)=
\nonumber \\
\frac{4\pi}{p_1}\Im\mbox{m}F^0_{11}(\bp_1,\bp_1)-\int d\hy\, |F^0_{11}(p_1\hy,\bp_1)|^2.
\label{sigma-a2}
\end{eqnarray}
This is the optical theorem for equation (\ref{DPsi-0}) and the annihilation cross section
$\sigma^a_2$
can be expressed now in terms of the left hand side as
\be
\sigma^a_2=\frac{2\nu_1(-g)}{\hbar^2 p_1}\int d\bx_1d\by_1\, {\Upsilon^{0+}_2}^*(\bp_1)
W_2\Psi^{0+}_2(\bp_1).
\label{sigma-a20}
\ee
To elucidate the further meaning of the cross section $\sigma^a_2$ it is instructive
to consider the limiting case as $g\to 0$.
In the limit no absorption potential is present, scattering becomes unitary and amplitude
$F^0_{11}$ coincides
with $f_{11}$ from (\ref{multichannelas1}). The standard unitary variant of the optical theorem for
the amplitude $f_{11}$ has the form
\be
\frac{4\pi}{p_1}\Im\mbox{m} f_{11}(\bp_1,\bp_1)-
\int d\hy\, |f_{11}(p_1\hy,\bp_1)|^2 - \int d\hy\, |f_{21}(p_1\hy,\bp_1)|^2=0
\label{OptTh-0}
\ee
where $f_{11}$ is the elastic $e^+-$H amplitude and $f_{21}$ is the rearrangement Ps$-p$ amplitude.
Therefore, the right hand
side of (\ref{sigma-a2}) has the limit
\be
\frac{4\pi}{p_1}\Im\mbox{m}F^0_{11}(\bp_1,\bp_1)-\int d\hy\, |F^0_{11}(p_1\hy,\bp_1)|^2 \to
\int d\hy\, |f_{21}(p_1\hy,\bp_1)|^2=\sigma_{21}.
\label{sigma-21}
\ee
The quantity  $\sigma_{21}$ is nothing but the positronium formation cross section.
At the same time the formula (\ref{sigma-a20})
leads to an uncertainty in the limit as $g\to 0$ when the diverging integral is multiplied
by the vanishing factor
$g$.
Actually, the resolution of this uncertainty is made by (\ref{sigma-21}) and gives the relation
\be
\sigma^a_2=\sigma_{21}+{\cal O}(g),
\label{sigma-a2-21}
\ee
which clearly shows that $\sigma^a_2$ is the cross section of the annihilation after the
positronium formation.

The main result of this subsection is the representation of the annihilation cross section
as the sum of two
terms
\begin{equation*}
\sigma^a=\sigma^a_1+\sigma^a_2,
\end{equation*}
where $\sigma^a_2$ is shown to represent
the cross section of the annihilation after the positronium formation,
which is given
by the formulae (\ref{sigma2}) or (\ref{sigma-a20}). Therefore $\sigma^a_1$ is the direct
annihilation cross section, given by (\ref{sigma1}).

The analysis made above shows that the definitions of cross sections in terms of scattering
amplitudes (\ref{sigma2},~\ref{sigma1}) are uniform and are valid below as well as above
the rearrangement
threshold of the  positronium formation. Moreover, by construction, the integrals (\ref{F-0-11},
\ref{F-1-11})
involved in the definitions of the amplitudes  $F^0_{11},\, F^1_{11}$
have the finite limit as $g\to 0$. This property will
be used below for perturbative calculations of the amplitudes and cross sections.
At the same time the integrals in representations (\ref{sigma_a},~\ref{sigma-a20}) are divergent
above the positronium formation threshold if the limit
$g\to 0$ is taken.
That makes these formulae not suitable for perturbative methods above the positronium formation
threshold. 

\subsection{Absorbing annihilation potential}
In this subsection we fix the coordinate form of the absorption potential by comparing the
definition
(\ref{sigma_a}) for $\sigma^a$ with QED formula for the $2\gamma$ singlet $e^+e^-$ annihilation
\footnote{The spin-averaging factor $1/4$ for singlet $2\gamma$ annihilation is implied
implicitly.}     \cite{fraser, charlton}
\begin{equation}
{\sigma}^{a}=\pi r^2_0 (c/v){Z}_{\mbox{{eff}}}.
\label{QED-sigma}
\end{equation}
In this formula $r_0$ is the classical electron radius, $c$ is the speed of light and $v$ is the incident
velocity of the positron.
The effective number of electrons ${Z}_{\mbox{{eff}}}$ participating in annihilation is given by the
integral
\begin{equation}
Z_{\mbox{eff}}= \int d{\bf x}_1 d{\bf y}_1\,
|\Psi^{0+}({\bf x}_1,{\bf y}_1)|^{2}
\delta({\bf x}_2).
\label{-Zeff}
\end{equation}
Here
$\Psi^{0+}$ is the $\Psi^{01}$ solution of the $e^+-$H scattering problem (\ref{Sch2},
\ref{multichannelas1}, \ref{multichannelas2})
when the absorption potential is not taken into account.
The integral in (\ref{-Zeff}) is well defined below the Ps-formation threshold. In this case (\ref{QED-sigma},
\ref{-Zeff}) can be considered as the first order perturbation approximation to (\ref{sigma_a}),
since below the rearrangement threshold 
$\Psi^+\simeq \Psi^{0+}$
is the well defined first order perturbation solution (Born)
to (\ref{DWLS}). This observation was used in \cite{mitroy}  to determine the
absorption potential
for $2\gamma$ singlet $e^+e^-$ annihilation as
\begin{eqnarray}
\im g W_2(\bx_2)=\im g \delta(\bx_2)
\label{delta-absorb}\\
g=-\frac{e^2}{a_0}2\pi \alpha^3.
\nonumber
\end{eqnarray}
Here $a_0$ is the Bohr radius and $\alpha$ is the fine structure constant. This potential was used in
\cite{mitroy}
to calculate the direct annihilation cross section by solving the Lippmann-Schwinger
equation for T-matrix
below the Ps-formation threshold.

The formula (\ref{-Zeff}) cannot be extended for calculations above the Ps-formation
threshold since the integral
diverges.
One of ways to go over the Ps-formation threshold is the use of (\ref{sigma_a})
with absorption potential
(\ref{delta-absorb}) incorporated into the Schr\"odinger equation. This was done in
papers \cite{iks, igarashi2} where the overall annihilation cross section $\sigma^a$ was computed
below as well as above the Ps-formation threshold within the hypersperical close coupling technique
for  the time independent three-body Schr\"odinger equation. Paper
\cite{yamanaka} represents the solution of the time-dependent three-body Schr\"odinger  equation with
the absorption potential (\ref{delta-absorb}) below as well as above the Ps-formation threshold.
 All these papers
dealt with  numerical solutions of respective equations and the delta-functional singularity
of the potential
(\ref{delta-absorb}) was treated by a certain numerical approximation. Nevertheless,
the analytical status of the
potential (\ref{delta-absorb}) is not satisfactory. The delta-functional singularity
is too strong and makes the Hamiltonian not well defined. This issue was not addressed in aforementioned
papers and we give a portion of the necessary analysis in this subsection.

It is well known already since the papers by Fermi \cite{delta-potF} and then Breit
\cite{delta-potB} that the
three-dimensional delta-function potential can be incorporated into
the Schr\"odinger equation only perturbatively.
One of the approaches to go beyond the perturbative treatment is the use of a zero-range potential
\cite{delta-potB, DemkovO}. There are two common ways to introduce the zero-range potential.
One is imposing boundary conditions for the wave-function
at the singularity point. The other one is introducing into the Hamiltonian an additional term,
which enforces the wave-function to fulfill  the boundary conditions. This  term
can conveniently be represented in the compact form by the quasi-potential \cite{BW, Huang}.
We choose the second option.

 The singularity caused by the zero-range potential in the case of the
electron-positron interaction is located
at the same point as the
Coulomb singularity $-e^2/x_2$. The latter leads to the modification of the standard
zero-range potential and of the respective quasi-potential. The resulting definition for
the coordinate part $W_2$ of the absorption potential is
\be
W_2(\bx)=\delta(\bx)\frac{1-n_2x}{1+n_2x\log x}\,\frac{d}{dx}\,\frac{x}{1+n_2x\log x}
\label{qdelta-pot}
\ee
where $n_2=-2\mu_2 e^2$ and $\mu_2$ is the $e^+e^-$ reduced mass.
The detailed derivation of (\ref{qdelta-pot}) involves a substantial portion of mathematics
and will be
published elsewhere. Some basic theorems, which define the general properties of the zero-range
potential
with the Coulomb modification, can be found in  \cite{Albeverio}.

It is straightforward to see that in the limit $n_2 \to 0$ the quasi-potential (\ref{qdelta-pot})
takes the
standard form
\be
W_2(\bx)\to \delta(\bx)\frac{d}{dx}\,x.
\ee
The quasi-potential $\im g W_{2}(\bx_2)$ enforces the following asymptotics for the wave function as
$x_2\to 0$
\be
\Psi(\bx_2,\by_2)\propto \frac{a(\by_2)}{4\pi}[\frac{1}{x_2}+n_2\log x_2] + b(\by_2) +
{\cal O}(x_2\log x_2),
\label{bc}
\ee
where $a/b=-\im g{2\mu_2}/{\hbar^2}$. This asymptotics, as usually,
determines the appropriate boundary conditions, which we do not write down here explicitly.
It can be shown that the action of the quasi-potential on the function with such an asymptotics
is given by the formula
\be
W_2(\bx_2)\Psi(\bx_2,\by_2)=  \delta(\bx_2)b(\by_2).
\ee
The latter means that the action of the quasi-potential on a function $\chi(\bx_2,\by_2)$,
which is regular in the point $x_2=0$, is equivalent to the delta-function
\be
 W_2(\bx_2)\chi(\bx_2,\by_2)=  \delta(\bx_2)\chi(0,\by_2).
\label{reg delta}
\ee
This formula shows that if a matrix element of the quasi-potential is calculated between the functions
$\chi,\,\omega$,
which are regular at  $x_2=0$, then the quasi-potential is equivalent to the delta-function, i.e.
\be
\langle \chi|W_2(\bx_2)|\omega\rangle = \langle \chi|\delta(\bx_2)|\omega \rangle .
\ee
This statement justifies the use of the delta-function as the coordinate part of the
absorption potential in \cite{iks, igarashi2, yamanaka},
 since the basis functions of the approaches used to compute the
matrix elements of the absorption potential are smooth. Nevertheless, any basis of smooth functions
cannot reproduce the singularity in (\ref{bc}) by a finite number of terms, what  always
happens  along the numerical solution. Hence, such a treatment of the zero-range potential in
\cite{iks, igarashi2, yamanaka} is approximative
but, in view of the fact $a(\by_2)=-\im g{2\mu_2}b(\by_2)/{\hbar^2}$ with $|g|\ll1$ for the $e^+e^-$ annihilation,
the approximation  is reasonable.
\section{Calculation of annihilation cross section}
The formalism developed above was applied for calculations of the annihilation in $e^+-$H collision
below the Ps-formation threshold and above the threshold in the  Ore gap. The latter
is defined as the interval of the energy between the Ps($1s$) and H($n=2$) thresholds.
All calculations were made on the basis of an extension of
the multichannel numerical algorithm for Faddeev equations, described in details in \cite{hu},
for the case of annihilation.
The algorithm uses the bipolar harmonic expansion to represent the angular dependence of the wave-function
components
\begin{eqnarray}
\Psi^+_i(\bx_i,\by_i)=\sum_{L l_1 l_2}\frac{\psi^L_{l_1 l_2}(x_i,y_i)}{x_iy_i}
{\cal Y}^{LM}_{l_1l_2}(\hx_i,\hy_i),
\label{Bipolexp}\\
{\cal Y}^{LM}_{l_1l_2}(\hx_i,\hy_i)=[Y^{m_1}_{l_1}(\hx_i)\otimes Y^{m_2}_{l_2}(\hy_i)]_{LM}.
\nonumber
\end{eqnarray}
This expansion reduces the MFE to a set of coupled equations for radial components
$\psi^L_{l_1,l_2}(x_i,y_i)$ which are then approximated by
the quintic-spline expansion and solved by the orthogonal collocation procedure. The maximum values
of $l_1, l_2$ used in (\ref{Bipolexp}) range from $12$ to $15$.

\begin{figure}
\begin{center}
{
\includegraphics[scale=0.7]{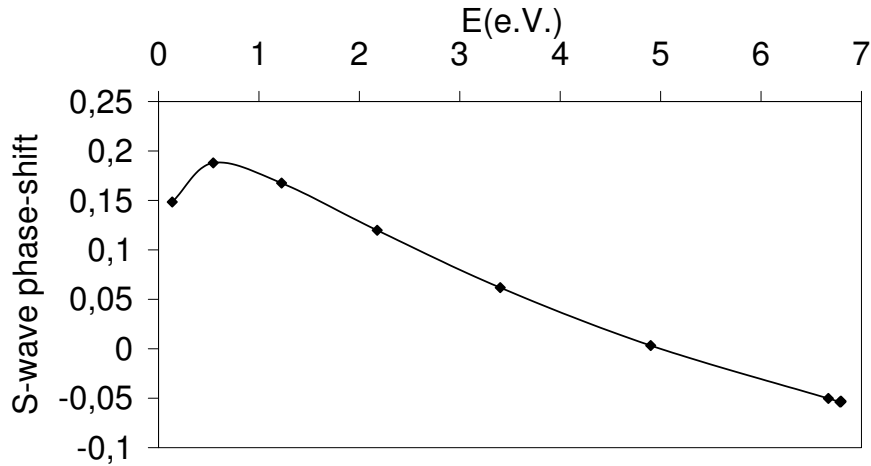}
}
\end{center}
\caption{$L=0$ phase-shift for $e^+-$H elastic scattering.}
\label{fig-delta0}
\end{figure}

In order to test the numerical approach, the genuine  Coulomb problem for $e^+-$H scattering was
solved on the platform of MFE (\ref{mfe-1}, \ref{mfe-2}) with $g=0$.
The results of calculations for s-wave $e^+-$H phase shift $\delta_0$
are given in  Table \ref{Tab1} together with data of other authors.
\begin{table}

\begin{center}
\begin{tabular}{rrrrrr
}\hline
$p_1$           & present work & \cite{Kvits-hu95}& \cite{Bhatia-T71} & \cite{Levin88}&\cite{Humb-Wallace72}
\\  \hline
0.1          & 0.1484  & 0.149  & 0.1483  & 0.152 & 0.148
\\ 
0.2          & 0.1879  & 0.189   & 0.1877   & 0.188 & 0.187
\\ 
0.3          & 0.1676  & 0.169  & 0.1677  & 0.166 & 0.167
\\ 
0.4          & 0.1198  & 0.121  & 0.1201  & 0.118 & 0.119
\\ 
0.5          & 0.0618 & 0.062  & 0.0624  & 0.061 & 0.062
\\
0.6          & 0.0032  & 0.003  & 0.0039  & 0.003 & 0.003
\\
0.7          & -0.0502  & -0.05  & -0.051  & 0.0053&
\\
\hline
\end{tabular}
\end{center}
\label{tab1}
\caption{$L=0$ phase shift $\delta_0$ for $e^+-$H elastic scattering.
Momenta $p_1$ are given in units of $a_0^{-1}$.
\label{Tab1}}
\end{table}
As one can see, the agreement of our calculations with previous results is quite good.
Figure \ref{fig-delta0}
provides an alternative representation of $\delta_0$ as the function of the energy to demonstrate the
regular character of the calculated phase shift $\delta_0$.
The calculations of the direct annihilation cross section $\sigma^a_1$ were performed
with the help of the
representation (\ref{sigma1}). The
amplitudes $F^0_{11}$ and $F^1_{11}$ were computed  from the solutions of the respective
Faddeev equations. The effective number $Z^1_{\mbox{eff}}$ for the direct annihilation
cross section was obtained using the standard expression
(\ref{QED-sigma})
\be
\sigma^a_1=\pi r^2_0 (c/v){Z}^1_{\mbox{{eff}}}.
\label{YH-sigma1}
\ee
As it was mentioned in   subsection \ref{OT1},
in the calculations of the amplitudes $F^0_{11}$ and $F^1_{11}$
we have systematically  approximated
the Faddeev components, involved in the matrix-elements for the amplitudes,
 by the solutions $\Psi^{00\pm}_i$ of the MFE~with~$g=0$. This statement can easily be justified
by the iterative  solution of the IMFE (\ref{IMFE-1}, \ref{IMFE-2}) for $|g|\ll 1$.
The iterative solutions is well defined thanks to the fact that the matrix kernel of the Faddeev equation
is proven to be compact. This is another advantage of the Faddeev three-body equations.
 As the result, the expression for the absorption amplitude $F^1_{11}$
from (\ref{F-1-11}) can be simplified as
\be
F^{1}_{11}(\bp'_1,\bp)=
\frac{-\nu_1}{2\pi\hbar^2}
\langle [\Psi^{00-}_1(\bp'_1)+\Psi^{00-}_2(\bp'_1)]|W_2|\Psi^{00+}_1(\bp_1)\rangle ,
\label{F-00-11}
\ee
which was actually used for calculations of this paper.

Since a number of data from other authors for direct annihilation cross section
below the Ps-formation threshold is available,
In Table \ref{Tab2} we display our calculated phase-shift $\delta_0$ and $Z^1_{\mbox{eff}}$
together with results of other authors obtained with the
standard formula (\ref{QED-sigma}) for one of the typical value of the relative
momentum $p_1=0.4\ [1/a_0]$.
\begin{table}
\begin{center}
\begin{tabular}{rrrrr}
\hline
Ref.             & $\delta_0$ & $Z_{\mbox{eff}}$\\ \hline
present work     & 0.11983  & 3.3293 \\ 
\cite{Brom-Mit03}& 0.1198   & 3.232  \\ 
\cite{Bhatia-T71}& 0.1201   & 3.327  \\ 
\cite{Grib-L03}  & 0.1198   & 3.407   \\ 
\cite{Mit-Rat95} & 0.1191   & 3.332
\\
\hline
\end{tabular}
\end{center}
\caption{$L=0$ phase shift $\delta_0$ for $e^+-$H elastic scattering and effective
number $Z_{\mbox{eff}}$ for the relative momentum $p_1=0.4\ [a_0^{-1}]$.
\label{Tab2}}
\end{table}
The agreement is very good for calculations made by quite different approaches.
In Table \ref{TabZ} we collect the results of existing calculations of $Z_{\mbox{eff}}$
below the Ps-formation threshold to compare with our results. One cannot expect the complete agreement
since our definition of $\sigma^a_1$ concerns  the direct process only, whereas
the standard definition below the Ps-formation threshold deals with the overall annihilation
cross section $\sigma^a$.
Nevertheless,  Table \ref{TabZ} shows that the difference is not so dramatic.
\begin{table}
\begin{center}
\begin{tabular}{rrrrr}
\hline
$p_1$     & present paper & \cite{Kvits-hu95} &\cite{Humb-Wallace72}& \cite{Bhatia-Drach74}\\
\hline
0.1       &7.2570          &7.55               &7.5                  &7.363 \\
0.2       &5.1627         &5.74               &5.7                  &5.538 \\
0.3       &4.1061         &4.36               &4.3                  &4.184 \\
0.4       &3.3293        &3.4                &3.3                  &3.327 \\
0.5       &2.8118         &2.74               &2.7                  &2.73  \\
0.6       &2.4625         &2.29               &2.3                  &2.279  \\
0.7       &2.2529         &2.02               &                     &1.950 \\
\hline
\end{tabular}
\end{center}
\caption{$L=0$  effective
number $Z_{\mbox{eff}}$ for $e^+-$H annihilation. The relative momenta $p_1$ are
given in units of $a_0^{-1}$.
\label{TabZ}}
\end{table}

\begin{figure}
\begin{center}
{
\includegraphics[scale=0.7]{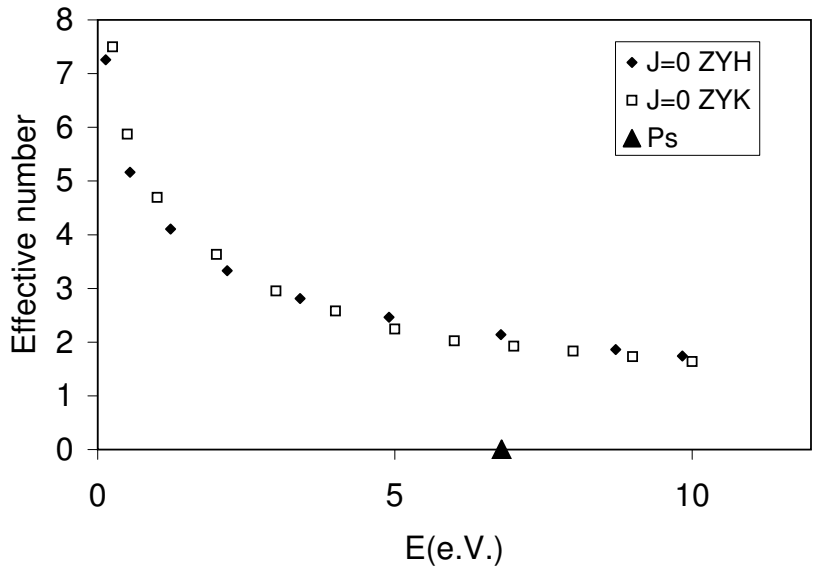}
}
\end{center}
\caption{The effective number $Z^1_{\mbox{eff}}$ for the $L=0$ direct annihilation.
The diamonds are the calculations of the present
paper, the open squares are the data from \cite{yamanaka}, the triangle marks the
positronium formation threshold.}
\label{fig2}
\end{figure}

The extensive calculations of the direct annihilation cross section $\sigma^a_1$ were made in the
interval of the energy between H($n=1$) and H($n=2$) thresholds on the basis of the formula
(\ref{sigma1}).
Figure \ref{fig2} shows the s-wave
effective number $Z^1_{\mbox{eff}}$ derived from $\sigma^a_1$,
with the formula (\ref{YH-sigma1}), and the results
of calculations from \cite{yamanaka}.
Although, quite different methods
(time-independent Faddeev equations and  time-dependent wave-packet approach to
the Schr\"odinger equation) of calculations
for the direct annihilation cross sections have been used, the agreement between data is fairly good.
In fact, the definition of the direct cross section $\sigma^a_1$ in our formalism as the remainder
of the overall
annihilation cross section $\sigma^a$ after subtraction of the annihilation cross section after the
Ps-formation
$\sigma^a_2$
\begin{equation}
\sigma^a_1=\sigma^a-\sigma^a_2,
\label{subtr}
\end{equation}
(see equation (\ref{sigma21})) is quite equivalent to the time-dependent definition by formula
(7) of ref. \cite{yamanaka}. Nevertheless, as the data of Tables \ref{Tabs}, \ref{Tabp}
shows,  the exact use of the formula (\ref{subtr})
would be very unpractical. The formation cross section $\sigma_{21}$, which is the leading term
of $\sigma^a_2$ due to (\ref{sigma-a2-21}), rapidly increases above the threshold and is several
order of magnitude bigger than $\sigma^a_1$.  The analytic separation of the overall annihilation
cross section into the formation and direct parts made in Subsection \ref{OT1} and given by formulae
(\ref{sigma21}-\ref{sigma1}) in terms of amplitudes $F^0_{11}$ and $F^1_{11}$
is therefore of the great practical importance.

\begin{figure}
\begin{center}
{
\includegraphics[scale=0.7]{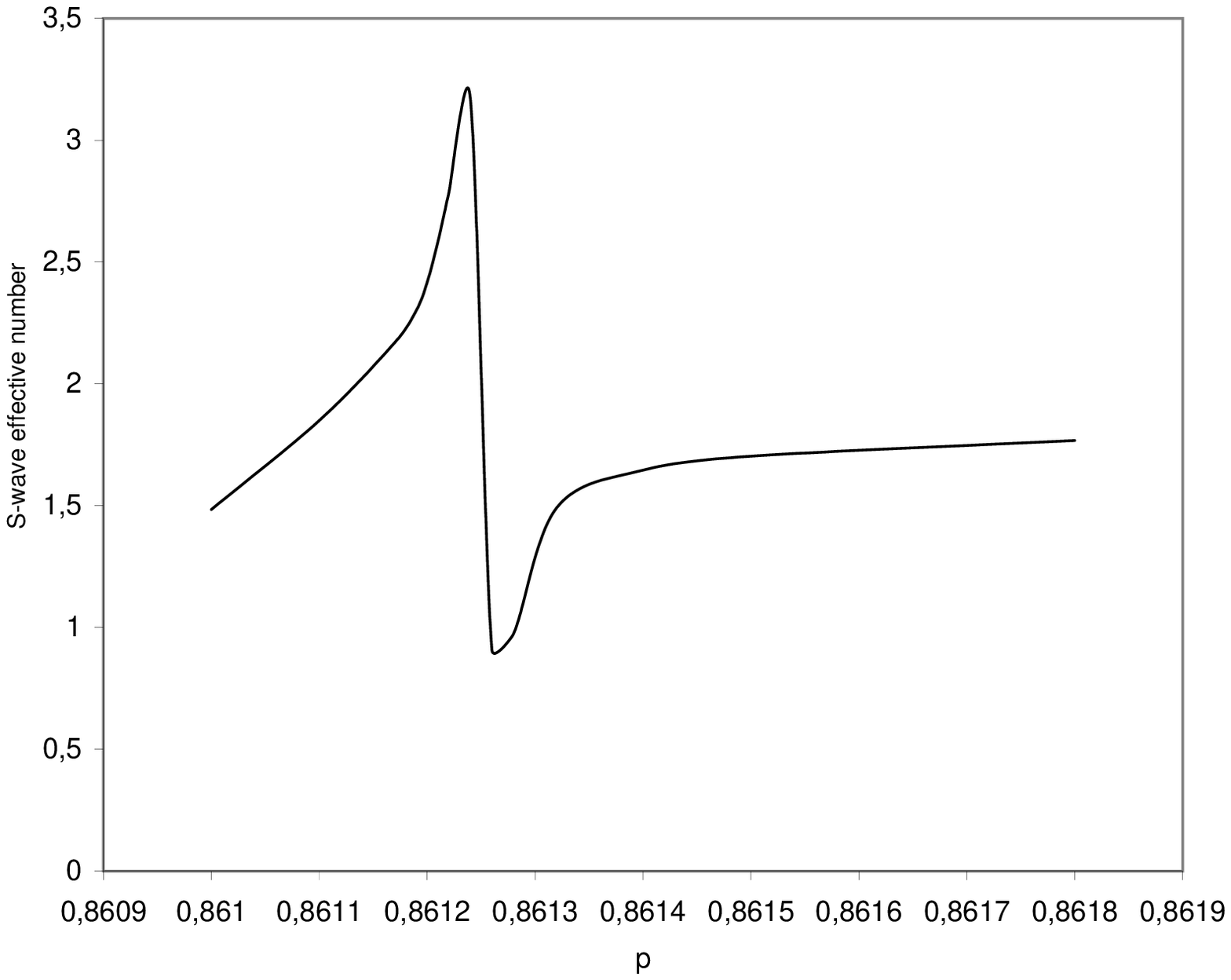}
}
\end{center}
\caption{$L=0$ effective number $Z^1_{\mbox{eff}}$ in the resonant region. The momenta $p$ are given in
$a_0^{-1}$ units.}
\label{a-res-s}
\end{figure}

\begin{figure}
\begin{center}
{
\includegraphics[scale=0.7]{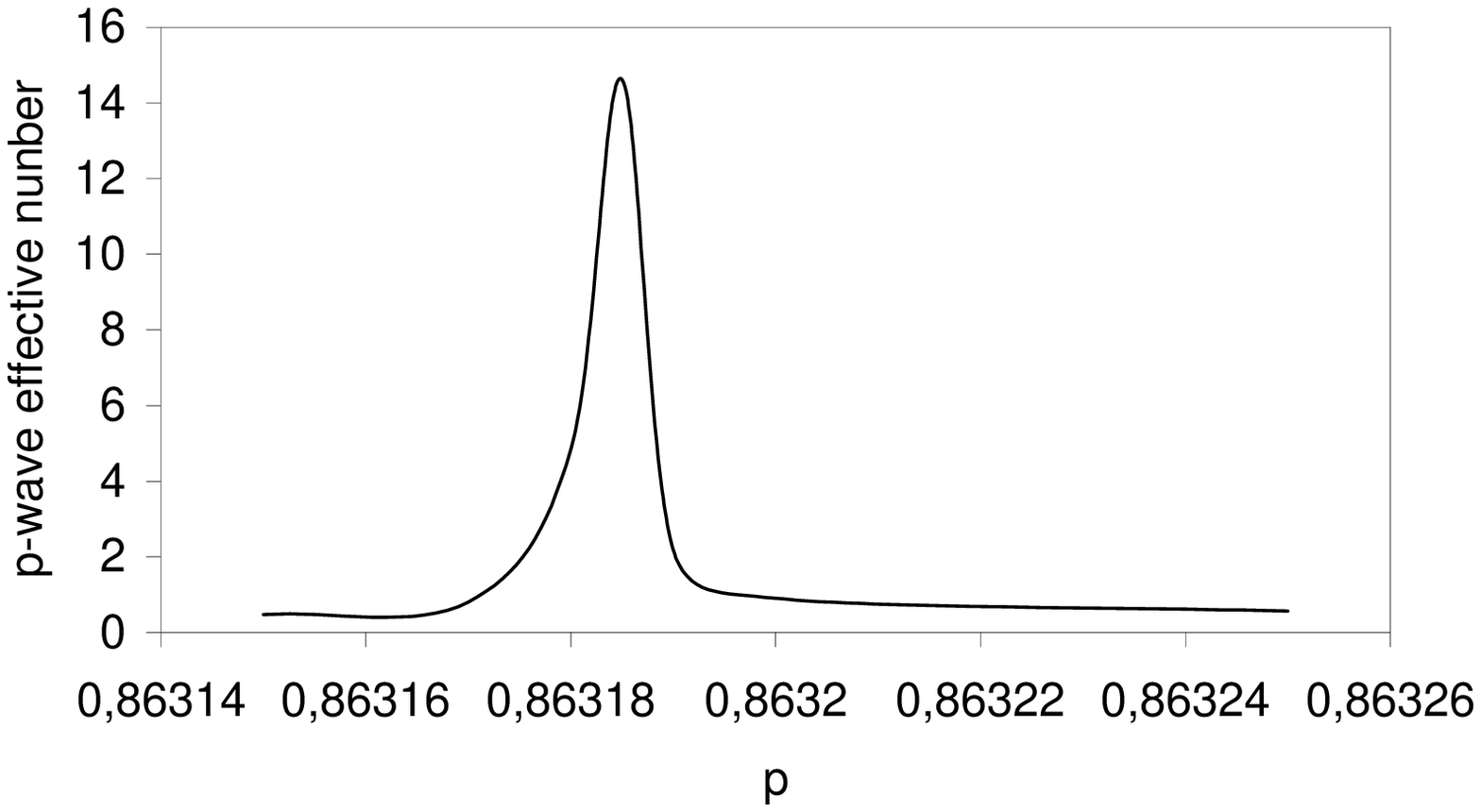}
}
\end{center}
\caption{$L=1$ effective number $Z^1_{\mbox{eff}}$ in the resonant region. The momenta $p$ are given in
$a_0^{-1}$ units.}
\label{a-res-p}
\end{figure}

In the following Tables \ref{Tabs} and \ref{Tabp} we present the results of our calculations for the direct
annihilation cross section together with the cross section of the positronium formation above the
Ps-formation threshold.
\begin{table}
\begin{center}
\begin{tabular}{rrrrr}
\hline
$p_1$       &  $Z^1_{\mbox{eff}}$ & $\sigma^a_1$ & $\sigma_{21}$ \\
\hline
0.70654 & 2.3289              & 1.28 [-6]     &  9.05 [-4]   \\
0.71    &  2.1715             & 1.19 [-6]     &  4.14 [-3]   \\
0.8     &  1.8640              & 9.05 [-7]     &  5.03 [-3]   \\
0.85    &  1.7404             & 7.95 [-7]    &5.83 [-3]    \\
0.861   &  1.4840              & 6.69 [-7]   &0.01087    \\
0.8611  & 1.9678              & 8.88 [-7]    &0.01682    \\
0.86118 &  2.2200               & 1.00 [-6]  &0.02694    \\
0.86119 & 2.3022              &1.04 [-6]     &0.02943 \\
0.8612  & 2.4175              &1.09 [-6]     &0.03247\\
0.86121 & 2.5832              &1.16 [-6]     &0.03622 \\
0.86122 & 2.7817              &1.25 [-6]     &0.04087\\
0.86124 & 3.1700                &1.43 [-6]   &0.05380 \\
0.86126 & 0.9735              &4.39 [-7]     &0.07030 \\
0.86128 & 0.9742              &4.39 [-7]     &0.07450 \\
0.86132 & 1.4915              &6.73 [-7]     &0.02089 \\
0.8614  & 1.6459              &7.42 [-7]     &0.00013 \\
0.8615  & 1.7034              &7.68 [-7]     &0.00083  \\
0.8618  & 1.7674              &7.97 [-7]     &0.00326 \\
\hline
\end{tabular}
\end{center}
\caption{$L=0$ effective number $Z^1_{\mbox{eff}}$, the direct annihilation cross section $\sigma^a_1$, and
the positronium formation cross section $\sigma_{21}$.
The cross sections are given in units of $\pi a_0^2$ and
momenta in units of $a_0^{-1}$. The abbreviation [-n] is used for $10^{-n}$.
\label{Tabs}}
\end{table}
\begin{table}
\begin{center}
\begin{tabular}{rrrrr}
\hline
$p_1$       &  $Z^1_{\mbox{eff}}$ & $\sigma^a_1$ & $\sigma_{21}$ \\
\hline
0.8      & 0.5404  &2.62 [-7] & 0.485\\
0.85    &0.6933  &3.17 [-7]    &0.566\\
0.8631  &0.4775  &2.15 [-7]    &0.749\\
0.86313 &0.4695  &2.11 [-7]    &0.848\\
0.86315 &0.4675  &2.10 [-7]    &1.022\\
0.86317 &0.5914  &2.66 [-7]    &1.772\\
0.86318 &4.8319  &2.17 [-6]    &3.680\\
0.863185 &14.6255 &6.58 [-6]    &1.770\\
0.86319 &2.1967  &9.88 [-7]    &0.129\\
0.8632  &0.9012  &4.06 [-7]    &0.068\\
0.86325 &0.5629  &2.53 [-7]    &0.407\\
\hline
\end{tabular}
\end{center}
\caption{$L=1$ effective number $Z^1_{\mbox{eff}}$, the direct annihilation cross section $\sigma^a_1$, and
the positronium formation cross section $\sigma_{21}$. The cross sections are given in units of $\pi a_0^2$ and
momenta in units of $a_0^{-1}$. The abbreviation [-n] is used for $10^{-n}$.
\label{Tabp}}
\end{table}
They are given for slightly different values
of momenta $p_1$ for $L=0$ and $L=1$  in order
to emphasize the most characteristic behavior of cross sections
near the respective Feshbach resonances.
 Besides the expected difference
in several order of magnitude between direct annihilation cross section $\sigma^a_1$ and
the positronium formation cross section $\sigma_{21}$, the strong correlation between
these cross sections in the region of the sharp increase of $\sigma^a_1$ across $0.86124\,
[a_0^{-1}]$ for
$L=0$ and $0.86318\, [a_0^{-1}]$ for $L=1$ is clearly seen.
The graphical representation of that sharp increase
of cross sections is given on Figures \ref{a-res-s} and \ref{a-res-p} which display
s- and p-wave $Z^1_{\mbox{eff}}$.
This resonant feature was also observed in the spatial behavior of the solution to the Faddeev equation.
The first component $\Psi^{+}_1$, which is related to the $e^+-$H channel,
always exhibits the characteristic  resonant bump near resonant energy.
These spatial resonant structures can be extensive depending on the lifetime of the resonances.
In \cite{hu-yak-papp}, reasonable estimation of the energy width of a number of Feshbach resonances
were made using the uncertainty principle  applied to the graphical representation
of the wave function components.

\section{Conclusion}
In present paper we have developed the multichannel time-independent formalism, which is capable
to describe the scattering and annihilation processes  in the positron-hydrogen collision above the
rearrangement threshold. The expression derived in the paper for the direct annihilation cross section
in terms of amplitudes
is proven to extend the standard formula to the energy region above the positronium formation threshold.
Below the threshold our cross section and the cross section calculated from the standard theory are
in good agreement provided the close vicinity of the positronium-formation threshold is not considered
where the standard cross section becomes infinite.
The direct annihilation cross section defined by the formula (\ref{sigma1}) does not exhibit
any singular behavior at the Ps-formation threshold. It is in good agreement with the nonsingular
direct annihilation cross section computed from the time-dependent solution of the three-body
Schr\"odinger equation for $e^+-$H system \cite{yamanaka}.

The formalism of this paper can readily be extended beyond the Ore gap.
 In this case the multichannel optical theorem, which  generalizes
(\ref{OptTh}), plays  the key role in the determination of cross sections.
Preliminary calculations indicated the much larger enhancement of the direct annihilation cross section
near the Feshbach resonances in the eight open channel region above the Ps($n=2$) threshold.
Since there are
numerous resonances beyond the Ore gap, they evidently made a significant contribution to the overall
annihilation peak around $12$ e.V. displayed in Figure 2 of \cite{yamanaka}.
We hope that our individual annihilation resonance structure will provide
a new experimental tool to study sharp resonances. Although at the present time, it is not
possible to conduct such an experiment for $e^+-$H system,
experiments for positron scattering on large molecules have been done for many systems
\cite{Surko}.

Our approach, in perspective,
opens the way to considering more complicated systems with more than three particles, as for example
$e^+e^+e^-e^-$.
Suitable formalism for multichannel scattering \cite{Merk-Yak-4},
which is the generalization of the Faddeev equations for the
four-particle systems,  will be helpful  as an important theoretical step towards the experimental
verification and utilization of the rich positron annihilation physics.

\ack
The authors appreciate the support of the NSF grant Phy-0243740, INTAS grant No. 03-51-4000 and the
generous supercomputer time awards from
grants MCA96T011 and TG-MCA96T011 under the NSF partnership for Advanced
Computational Infrastructure,
Distributed Terascale Facility (DTF) to the Extensible Terascale Facility.
In  particular we are thankful to PSC and SDSC. The authors would like to thank
Dr. Y.~Kino~ for providing us with numerical data corresponding to the Figure 2 of
ref. \cite{yamanaka}, which made the comparison of our results much feasible. We are
thankful to Prof. K.A. Makarov for fruitful discussion on zero-range potentials.


\section*{References}
\begin{thebibliography}{20}
%
\bibitem{rich} Rich~A 1981 {\it Rev.\ Mod.\ Phys.} {\bf 53} 127
\bibitem{fraser} Fraser~P~A 1968 {\it Adv.\ At.\ Mol.\ Phys.} {\bf 4} 63
\bibitem{charlton} Charlton~M, Humberstone~J~W 2001
{\it Positron Physics}\ (Cambridge, UK: Cambridge University Press)
\bibitem{mitroy} Ivanov~I~A, Mitroy~J 2000 {\it J.\ Phys.\ B:\ At.\ Mol.\ Opt.
\ Phys.}  {\bf 33} L831
\bibitem{iks}
Igarashi~A, Kimura~M, Shimamura~I 2002
{\it Phys.\ Rev.\ Lett.}  {\bf 89} 123201
\bibitem{gl} Gribakin~G~F,  Ludlow~J 2002
{\it Phys.\ Rev.\ Lett.}  {\bf 88} 163202
\bibitem{igarashi2}
Igarashi~A, Kimura~M, Shimamura~I, Toshima~N  2003 {\it Phys.
Rev. A}  {\bf 68} 042716
\bibitem{yamanaka} Yamanaka~N, Kino~Y, Takano~Y, Kudo~H 2003
{\it Phys. Rev.} A {\bf 67} 052712
\bibitem{delta-potF} Fermi~E 1936 {\it Ricerca Sci.} {\bf 7} 13
\bibitem{delta-potB} Breit~G 1947 {\it Phys. Rev.} {\bf 71} 215
\bibitem{DemkovO} Demkov~Yu~N,  Ostrovskii~V~N 1988
{\it Zero-Range Potentials and their Applications in Atomic Physics} (New York: Plenum)
\bibitem{Laricchia} Van Reeth P, Laricchia G, Humberston J W 2005
{\it Phys. Scripta} {\bf 71} C9-13
\bibitem{hu-yak-papp} Hu~C-Y, Yakovlev~S~L, Papp~Z 2006 {\it Nucl. Instr. Meth.} {\bf B 247} 25
\bibitem{Messia} Messia~A  1958 {\it Quantum Mechanics} (New-York: J Wiley and Sons, Inc. )
\bibitem{Tobocman} Foldy~L~L, Tobocman~W 1957 {\it Phys. Rev.} {\bf 105} 1099
\bibitem{Newton}  Newton~R~G 1982 {\it Scattering
Theory of Waves and Particles} (New-York: Springer-Verlag New-York Inc.)
\bibitem{Schmid} Schmid~E~W, Ziegelman H  1974
{\it The quantum mechanical three-body problem} (Braunschweig: Vieweg)
\bibitem{Fadd} Faddeev~L~D  1961 {\it Sov. Phys. JETP} {\bf 12} 1014\\
Faddeev~L~D, Merkuriev~S~P 1993 {\it Quantum Scattering
Theory for Several Particle Systems}\  (Dordrech: Kluver)
\bibitem{merkuriev} Merkuriev S~P~ {\it Ann.~Phys.~(NY)}  {\bf 130}  395
\bibitem{yak-evans-hof} Yakovlev S~L~  1996 {\it Theor. Math. Phys.} {\bf 107}  835\\
Yakovlev S~L~1999 {\it Few Body Systems Supplement} {\bf 10} 85\\
Evans~J~W 1981 {\it J. Math. Phys.} {\bf 22} 1672\\
Evans~J~W, Hoffman~D~K 1981 {\it J. Math. Phys.} {\bf 22} 2858
\bibitem{Messia-was} Messia~A  1958 {\it Quantum Mechanics} (New-York: J Wiley and Sons, Inc.),
see CH. XIX, \S 2 Eq. XIX.14
\bibitem{BW} Blatt~J~M, Weisskopf~V~F 1952 {\it Theoretical Nuclear Physics}
(New-York: J Willey and Sons, Inc.)
\bibitem{Huang} Huang~K, Yang~C~N 1957 {\it Phys. Rev.} {\bf 105} 767
\bibitem{Albeverio} Albeverio~S, Gesztesy~F, H{\o}egh-Krohn~R, Holden~H 2005
{\it Solvable Models in Quantum Mechanics} (Providence, Rhode Island: AMS Chelsea Publishing)
\bibitem{hu} Hu~C-Y~ 1999  {\it J.\ Phys.\ B: At.\ Mol.\ Opt.\ Phys.\ } {\bf 32}, 3077
\bibitem{Kvits-hu95} Kvitsinsky A~A, Wu A, Hu~C-Y~1995~
{\it J.\ Phys.\ B: At.\ Mol.\ Opt.\ Phys.\ }~{\bf 28} 275
\bibitem{Bhatia-T71} Bhatia~A~K, Temkin~A, Drachman~R~J, Eiserike~H
~1971~{\it Phys. Rev. A}~{\bf 3}~1328
\bibitem{Levin88} Levin~F~S, Shetzer~J~ 1988~{\it Phys. Rev. Lett.}
~{\bf 61}~1089
\bibitem{Brom-Mit03} Bromley~M~W~J,  Mitroy~J~2003~{\it Phys. Rev. A} {\bf 67} 062709
\bibitem{Grib-L03} Gribakin~G~F,  Ludlow~J~2003~{\it Phys. Rev. A}~ {\bf 70}~ 032720\\
Van~Reeth~P,  Humberston~J~W~1997~{\it J. Phys. B:\ At.\ Mol.\ Opt.
\ Phys.} {\bf 30} 2477\\
Van Reeth~P,  Humberston~J~W~1998~{\it J. Phys. B:\ At.\ Mol.\ Opt.
\ Phys.} {\bf 31} L231
\bibitem{Mit-Rat95}  Mitroy~J,  Ratnavelu~K~ 1995~{\it J. Phys. B:\ At.\ Mol.\ Opt.
\ Phys.} {\bf 28} 287\\
 Ryzhikh~G~G,  Mitroy~J~2000~{\it J. Phys. B:\ At.\ Mol.\ Opt.
\ Phys.} {\bf 33} 2229
\bibitem{Humb-Wallace72}Humberstone J~W, Wallace J~B~G~1972~
{\it J. Phys. B: At. Mol. Phys.} {\bf 5}  1138
\bibitem{Bhatia-Drach74}Bhatia~A~K, Drachman~R~J, Temkin~A~ 1974~{\it Phys. Rev. A} {\it 9}  223
\bibitem{Surko} Barnes L~D,  Gilbert S~J,  Surko C~M  2003 {\it
     Phys. Rev. A} {\bf 67}, 032706\\
     Surko C~M, Gribakin G~F, Buckman S~J 2005 {\it J. Phys. B: At. Mol. Opt. Phys.} {\bf 38} R1-R70\\
     Barnes L~D, Young J~A, Surko C~M 2006 {\it Phys. Rev. A} {\bf 74} 012706



\bibitem{Merk-Yak-4} Merkuriev~S~P, Yakovlev~S~L~1982~{\it
Doklady AN USSR}~  {\bf 262} No. 3, 591
\\
Merkuriev S~P, Yakovlev~S~L, Gignoux~C~1984 {\it
Nucl. Phys.} {\bf A431} 125


\end{thebibliography}
\end{document}